\documentclass{epl}
\usepackage{amsmath}
\usepackage{amsfonts}
\usepackage{amssymb}
\usepackage{epsfig}
\usepackage{psfrag}

\title{Local dielectric permittivity near an interface}
\author{V. Ballenegger \and J.-P. Hansen}
\institute{
  Department of Chemistry, University of Cambridge, Cambridge CB2 1EW, UK
}
\shortauthor{V. Ballenegger \etal}
\pacs{61.20.Gy}{Theory and models of liquid structure}
\pacs{77.22.Ch}{Permittivity (dielectric function)}

\newcommand{\eq}[1]{\begin{equation} #1 \end{equation}}
\newcommand{\sr}{_{\mathrm{sr}}}
\newcommand{\lr}{_{\mathrm{lr}}}
\newcommand{\philr}{\phi^{\mathrm{lr}}}

\newcommand{\ext}{_{\mathrm{ext}}}

\newcommand{\dd}{_{\mathrm{dd}}}
\newcommand{\dc}{_{\mathrm{dc}}}
\newcommand{\cd}{_{\mathrm{cd}}}
\newcommand{\cc}{_{\mathrm{cc}}}

\newcommand{\be}{\beta}
\newcommand{\eps}{\epsilon}
\newcommand{\vp}{\vect{p}}
\newcommand{\hvp}{\hat{\vect{p}}}
\newcommand{\hve}{\hat{\vect{e}}}
\newcommand{\ver}{\vect{r}}
\newcommand{\E}{\vect{E}}
\newcommand{\vP}{\vect{P}}
\newcommand{\R}{\vect{R}}
\newcommand{\grad}{\vect{\nabla}}

\newcommand{\romd}{\upd}

\newcommand{\Graph}[1]{ \raisebox{-0.35 \height}{\epsfig{file=#1}} }

\begin{document}
\newcommand{\reg}{_\sigma}

\maketitle

\begin{abstract}
A statistical mechanics expression is derived for the space-dependent dielectric permittivity of a polar solvent near an interface. The asymptotic behaviour of this local permittivity is calculated in the low density limit, near a planar interface. The potential of mean force between two ions is shown to agree with the prediction of macroscopic electrostatics for large separations parallel to the interface.
\end{abstract}

\section{Introduction}
An efficient statistical description of complex fluids and biomolecular assemblies generally requires some degree of coarse-graining to be tractable. A good example is the ``implicit solvent'' description of colloids or biomolecules in aqueous solution where water is replaced by a polarizable continuum. An important, largely unsolved, question that arises is that of inhomogeneous screening of solute ions near interfaces. Can the reduction of the bare Coulomb interactions between charges be described by a spatially varying permittivity $\eps(\ver)$, and can one establish a rigorous statistical mechanics foundation of such a local permittivity, similar to Kirkwood's famous expression~\cite{Kirk} relating the constant permittivity $\eps$ of a bulk fluid to dipolar fluctuations? While many phenomenological prescriptions have been put forward for the determination of local permittivities~\cite{Schaefer}, the second question is so far largely unanswered. In a recent paper~\cite{Reimar}, an attempt was made to calculate the permittivity of a polar fluid near a planar interface by a generalization of Onsager's cavity model~\cite{Onsager}. In this letter, we reconsider the problem in a general statistical mechanics framework and a number of exact results are established to assess the validity of the concept of a local permittivity.

\section{Molecular model}
We consider a classical polar fluid made up of $N$ identical molecules confined to a volume $V$, at temperature $T$. The surrounding media that confine the fluid may be polarisable, and are treated using macroscopic electrostatics. The molecules carry a permanent dipole moment $\vp$ and are assumed to be non polarisable.
The position and the dipole moment of the $i$th molecule are denoted collectively by $i = (\ver_i,\vp_i)$.
The molecules interact via the pair potential
\eq{	\label{def v(1,2)}
    V(1,2) = v\sr(1,2) + \Theta(r_{12}-\sigma)\, (\vp_1 \cdot \grad_1) (\vp_2 \cdot \grad_2) G(\ver_1,\ver_2), \qquad \ver_{12}=\ver_2-\ver_1,
}
where the first term is a short-range interaction\footnote{The potential $v\sr(1,2)$ must decay at least as $|\ver_{12}|^{-4}$ at large relative distances. In addition to short-range repulsion, it may include the effects of higher multipole moments of the molecules.}
and the second term is the dipolar interaction with a cut-off for distances $r_{12}=|\ver_{12}|$ less than the molecular diameter~$\sigma$ [$\Theta$~is the Heaviside function]. The Green's function $G(\ver_1,\ver_2)$ is the solution of the equation $\Delta G(\ver_1,\ver_2)=-4\pi\delta(\ver_2)$ for $\ver_1,\ver_2 \in V$, with the appropriate boundary conditions dictated by the surrounding media. In the case of a planar interface between a continuous dielectric medium of dielectric constant $\eps'$ in region $z<0$ and a polar fluid in region $z>0$ (see Fig.~1), $G$ is given by
\eq{	\label{G interface}
   G(\ver_1,\ver_2) = \frac{1}{|\ver_{1}-\ver_2|} + \frac{1-\eps'}{1+\eps'}\, \frac{1}{|\ver_{1}-\ver_2^*|}, \qquad \ver^*_{2} = \ver_2-2(\ver_2\cdot\hve_z)\hve_z,
}
where the second term corresponds to the interaction with the image charge located at $\ver^*_{2}$. The presence of the surrounding polarisable media gives also rise to the one-body potential
\eq{	\label{def U(1)}
  U(1) = \frac{1}{2}\lim_{\ver_2\to\ver_1} (\vp_1 \cdot \grad_1) (\vp_1 \cdot \grad_2)
  	 \left[ G(\ver_1,\ver_2) - \frac{1}{|\ver_1-\ver_2|} \right]
}
which corresponds to the interaction of a dipole with its own image(s).\footnote{The factor $\frac{1}{2}$ occurs because, in a charging process where the dipole moments of the molecules are increased from zero to their final value, the infinitesimal energy increment $\romd U(1)$ is a quadratic function of $|\vp_1|$.}

\begin{figure}[h]
\psfrag{e}{{\Large $\eps'$}}
\psfrag{0}{0}
\psfrag{z}{$\hve_z$}
\onefigure{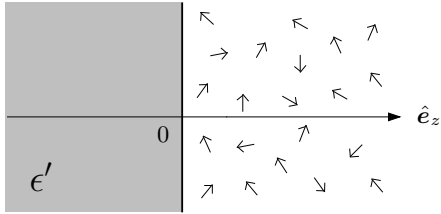}
\caption{A polar fluid confined to the region $z>0$.}
\label{f.1}
\end{figure}



\section{Polarisation in the linear regime}
When an external electric field $\vect{E}_0(\vect{r})=-\vect{\nabla}{\phi}(\vect{r})$ is applied to the fluid\footnote{$\vect{E}_0(\vect{r})$ is by definition the electric field in the absence of the fluid (the subscript 0 stands for 'no molecules').}, the latter acquires a non vanishing mean polarisation given by 
\begin{equation}
\label{e.1}
	\vect{P}(\vect{r}) = \int  \rho_{\phi}(\vect{r},\vect{p}) \, \vect{p} \, \upd\Omega 
\end{equation}
where $\rho_{\phi}(\vect{r},\vect{p})$ is the one-body density. Since the coupling energy of the $i{}^{\mathrm{th}}$ molecule to the external field is $\psi(i) = -\vect{p}_i\cdot\vect{E}_0(\vect{r}_i)$, the variation $\delta\rho(1) \equiv \rho_\phi(1)-\rho_{\phi=0}(1)$ of the one-body density due to the perturbation is given in linear response by
\begin{equation}
\label{e.2}
	\delta\rho(1) = -\beta \psi(1) \rho(1) + \rho(1) \int\upd 2\, h(1,2)
		\, \rho(2) (-\beta \psi(2))
\end{equation}
where $h(1,2)$ is the pair correlation function of the unperturbed system. Using the Ornstein-Zernike equation, we see that $\delta\rho(1)$ is linked to the direct correlation function via
\begin{equation}
\label{LMWB}
	\delta\rho(1) = -\beta \psi(1) \rho(1) + \rho(1) \int\upd 2\, c(1,2) \, \delta\rho(2)
\end{equation}
In a polar fluid, we expect the direct correlation function to decay at large distances like the dipolar potential. The integral in eq.~\eqref{LMWB} depends thus in general on the shape $\partial V$ of the sample. Following Ramshaw~\cite{Ramshaw}, let $c(1,2) = c\sr(1,2) + c\dd(1,2)$ where $c\sr(1,2)$ is a short-ranged function and $c\dd(1,2)=-\beta v\dd(1,2)$ where
\eq{	\label{def c_dip}
    v\dd(1,2) =  \Theta(r_{12}-\lambda)\, (\vp_1 \cdot \grad_1) (\vp_2 \cdot \grad_2) G(\ver_1,\ver_2) + \frac{4\pi}{3} \vp_1\cdot \vp_2\, \delta(\ver_{12}),
    \qquad \!\!\! \lambda\to 0,
}
is the electrostatic interaction between two ideal dipoles. The definition of $c\dd(1,2)$ at short range is of course arbitrary, and the delta function, which appears naturally in the definition of the dipolar potential (see~\cite{Jackson}), is included here for convenience. The cutoff $\Theta(r_{12}-\lambda)$ excludes the singularity of the dipolar potential at the origin. Integrals of $v\dd(1,2)$ are hence well defined (for finite volumes $V$), and the limit $\lambda \to 0$ is taken at the end of the calculation. Using \eqref{e.1} and \eqref{def c_dip}, we obtain
\eq{	\label{eq int E}
   \delta\rho(1) = \rho(1) \be \vp_1\cdot \E(\ver_1) + \rho(1) \int\romd 2\,
      c\sr(1,2)\,\delta\rho(2)
}
where $\E(\ver_1) \equiv \E_0(\ver_1) -  \grad_1 \int  \,  
      (\vP(\ver_2)\cdot\grad_2) G(\ver_1,\ver_2)
      \,  \romd\ver_2$ is the average electric field at $\ver_1$. Notice that the macroscopic field [and not the Lorentz field $\E_{\mathrm{L}}(\ver)=\E(\ver)+4\pi\vP(\ver)/3$] appears in eq.~\eqref{eq int E} because the delta function term provides exactly the ``missing'' contribution from the small excluded volume of radius $\lambda$. The shape dependence is now entirely embodied in the Maxwell field $\E(\ver_1)$ since the integrand in eq.~\eqref{eq int E} decays rapidly. By iterating eq.~\eqref{eq int E} and substituting into eq.~\eqref{e.1}, we find
\eq{	\label{P int E}
   \vP(\ver_1) = \rho(1) \be \vp_1\cdot \E(\ver_1) + \beta\int\romd\Omega_1\int\romd 2\,
    \rho(1)\rho(2) h\sr(1,2)\,\vp_1 (\vp_2\cdot\E(\ver_2))
}
with
\eq{	\label{OZ h_sr}
   h\sr(1,2) = c\sr(1,2) + \int\romd 3\, c\sr(1,3) \rho(3) h\sr(3,2).
}
Equation \eqref{P int E} relates the polarisation to the average electric field via an integral kernel. A local permittivity tensor $\vect{\eps}(\ver)$ can be defined when this relation is local. In that case,
\eq{	\label{P prop E}
   \vP(\ver) = \frac{\vect{\eps}(\ver)-\mathbf{1}}{4\pi} \E(\ver)
}
with
\eq{	\label{def eps_ij}
  \eps_{ij}(\ver_1) \equiv \delta_{ij} + 4\pi \beta \int\!\romd\ver_2\int\!\romd\Omega_1\romd\Omega_2\,
  (\vp_1\cdot \hve_i)(\vp_2\cdot \hve_j) \rho(1)[\delta(1,2) + \rho(2) h\sr(1,2)].
}
This is the general statistical mechanical expression of a local dielectric permittivity $\vect{\eps}(\ver)$. For points $\ver_1$ far from the boundaries, the fluid is homogeneous and isotropic [$\rho(1)=\rho/(4\pi)$], and eq.~\eqref{def eps_ij} reduces to the bulk dielectric constant of the fluid:
\eq{	\label{eps bulk}
   \frac{\eps - 1}{3y} =  \frac{1}{4\pi} \int\!\romd \ver_2\int\romd\Omega_1\romd\Omega_2 \, \hvp_1\cdot\hvp_2 \left[\delta(1,2) + \frac{\rho}{4\pi}h\sr^{\text{bulk}}(1,2)  \right]
   , \qquad y=\frac{4\pi \beta \rho p^2}{9}.
}
The present calculation follows the standard approach \cite{reviews} to the molecular theory of the dielectric constant based on the decomposition of the direct correlation function into a short-range and a dipolar part.\footnote{A different approach has been proposed recently by Alastuey and Ballenegger~\cite{AlaBal}.} The condition that the electric field must not vary appreciably over a distance of the order of the range of $h\sr(1,2)$ had already been noted in the pionneering work of Nienhuis and Deutch~\cite{ND}. We stress that eq.~\eqref{def eps_ij} does not express the permittivity in terms of local fluctuations of the dipole moment, because $h\sr(1,2)$ is only \emph{a part} of the total correlation function. Whether such a relation can be established in the general case of a local permittivity tensor $\vect{\eps}(\ver)$ is under current study.

Let us apply the result \eqref{def eps_ij} to the situation illustrated in Fig.~1. The symmetry of the planar interface implies that $\vect{\eps}(\ver)=\vect{\eps}(z)$ is diagonal, with $\eps_{xx}=\eps_{yy}=\eps^{\parallel}$ and $\eps_{zz}=\eps^{\perp}$. When the external field $\E_0=E^\parallel \hat{\vect{e}}_x$ is parallel to the interface, $\grad \wedge \E(\ver)=0$ implies $\frac{\partial}{\partial z}E^\parallel(z)=0$, and hence $\E(z)=E^\parallel \hat{\vect{e}}_x$ is constant throughout space. In this particular case, eq.~\eqref{P prop E} follows exactly from eq.~\eqref{P int E}. Very close to the interface, the permittivity $\eps^\parallel(z)$ is expected to vary on a microscopic scale, and the concept of a local permittivity becomes really meaningful only for sufficiently large $z$, where $\eps^\parallel(z)$ varies over mesoscopic distances.

In the case of a perpendicular external field $\E\ext=E^\perp \hat{\vect{e}}_z$, the Maxwell field  $\E(z)$ displays a non trivial profile along the $z$-axis. The local permittivity $\eps^\perp(z)$ is therefore well defined only in regions sufficiently far from the interface, where the variations of $\vect{E}(z)$ are neglegible over the range of decay of the correlation function $h\sr(1,2)$.

The leading deviations of the local permittivity $\eps_{ij}(z)$ from the bulk dielectric constant~$\eps$ of the fluid can be calculated explicitely for large $z$ and low densities. For a planar interface, the interaction $U(1)$ [eq.~\eqref{def U(1)}] of a dipole $\vp_1$ with its own image is given by
\eq{
   U(1) = \frac{p^2}{2}\frac{1-\eps'}{1+\eps'} \frac{1+\cos^2\theta_1}{(2 z_1)^3},
   \qquad \cos\theta_1=\hvp_1\cdot\hve_z.
}
We consider first the case $\eps'\neq 1$. The one-body density is then given at low density by $\rho(1)\simeq \rho/(4\pi)[1-\beta U(1)]$ when $z_1\to\infty$. Substituting in eq.~\eqref{def eps_ij} and neglecting all terms of order $\rho^2$, we see that the presence of the interface adds to Debye's well known low-density result $\eps\simeq 1+3 y$ the following surface corrections
\begin{equation}	\label{e.14}
   \eps^{\parallel}(z) \simeq \eps - \frac{1-\eps'}{1+\eps'} \frac{9y}{40}\frac{\be p^2}{z^3},
   \qquad
   \eps^{\perp}(z) \simeq \eps -  \frac{1-\eps'}{1+\eps'}\frac{33y}{80} \frac{\be p^2}{z^3}.
\end{equation}
When $\eps'=1$, $U(1)$ vanishes and there is no surface term at order~$\rho$. The correction of order $\rho^2$ can be obtained exactly by expanding $\rho(1)$ at low density. Treating the wall as a giant impenetrable particle (labelled 'w'), we have $\rho(1)=\rho^{(2)}(1,w)/\rho(w)$. The two-body density $\rho^{(2)}(1,w)$ can then be expanded using Mayer graphs. At order $\rho^2$, a single graph contributes to $\rho(1)$ when $z_1 \to \infty$, and we find (see~\cite{Bad1} for the details of this calculation)
\eq{	\label{e.15}
   \rho(1) \simeq \frac{\rho}{4\pi}
   \left[ 1-\frac{3y}{2} \beta p^2  \frac{1+\cos^2\theta_1}{16  z_1^3} + {\cal O}(\rho^2) \right],
   \qquad z_1\to\infty.
}
Notice that the factor $3y/2$ can be interpreted as the image charge factor $(\eps-1)/(\eps+1)$, expected from a macroscopic electrostatics calculation, when $\eps\to 1+3y$. Inserting eq.~\eqref{e.15} into \eqref{def eps_ij} shows that the deviations of $\eps_{ij}(z)$ to the bulk $\eps$ are given in the case $\eps'=1$ (up to order $\rho^2$ included) by the same expressions \eqref{e.14} with additional factors $3y/2$.

The above low-density results agree with the asymptotic expression of $\vect{\eps}(z)$ obtained in~\cite{Reimar} using a generalization of the Onsager cavity model to the case of a planar interface. An exact asymptotic expression for $\rho(1)$ at large distances for \emph{finite} densities has been obtained by Badiali~\cite{Bad1} and Zhang, Badiali and Su~\cite{ZhangBadSu} using Mayer diagrammatic techniques. Their general expression for $\rho(1)$ involves a term with an integral over the three-body direct correlation function $c(1,2,3)$. A similar analysis applied to $\eps_{ij}(z)$ shows that this function decays asymptotically as $z_1^{-3}$, with a coefficient that involves integrals over two, three and four-body direct correlation functions.


\section{Potential of mean force} The screening properties of the polar fluid close to the interface can be probed by introducing two fixed ions of charge $q_1$ and $q_2$ at $\ver_1=(\R_1,z_1)$ and $\ver_2=(\R_2,z_2)$, with $z_1,z_2>0$. Our aim is to determine the exact long-range behaviour of the potential of mean force $w(1,2)=w(q_1,q_2,z_1,z_2,R_{12})$ between the ions when their relative distance $R_{12}$ along the wall is very large. From \eqref{G interface}, the Coulomb interaction energy between the two ions is
\eq{
   V\cc(1,2) = q_1 q_2 G(\ver_1,\ver_2)=\frac{q_1 q_2}{|\ver_{12}|} + \frac{1-\eps'}{1+\eps'}\, \frac{q_1 q_2}{|\ver^*_{12}|}.
}
The interaction energy $V\cd(i,j)= V\dc(j,i)$ between an ion $i=1,2$ and a dipole $j$ has a short-range part~$v\cd^{\mathrm{sr}}(i,j)$ and a long-range part
\eq{	\label{def v_cd}
    v\cd^{\mathrm{lr}}(i,j)= v\cd(i,j)+ \frac{1-\eps'}{1+\eps'}\, v\cd^*(i,j),
    \qquad v\cd(i,j) = - (\vp_j\cdot\grad_j)\frac{q_i}{|\ver_{ij}|}.
}
The image interaction $v\cd^*(i,j)$ is obtained from $v\cd(i,j)$ by replacing $\ver_{ij}$ by $\ver^*_{ij}$. Similarly, the interaction energy $V\dd(i,j)$ between two solvent molecules has also a short-range part $v\dd^{\mathrm{sr}}(i,j)$ and a long-range part
\eq{
   v\dd^{\mathrm{lr}}(i,j) = v\dd(i,j) +  \frac{1-\eps'}{1+\eps'}\, v\dd^*(i,j)
}
where $v\dd(1,2)$ is defined in \eqref{def c_dip} and $v\dd^*(i,j)$ is the image dipole interaction.

The potential of mean force $-\beta w(1,2)$ is given formally by $-\beta V\cc(1,2)$ plus the sum of all connected Mayer graphs made of two root points representing the ions, internal points of weight $\rho(i)=\rho(z_i,\vp_i)$ representing solvent molecules, and Mayer bonds $f$ which take the value $f\dd(i,j)=\exp[-\beta V\dd(i,j)]-1$ between two solvent molecules and $f\cd(i,j)=\exp[-\beta V\cd(i,j)]-1$ between an ion $i$ and a dipole $j$ (see for example~\cite{Attard}). The graphs must be free of articulation points, and the root points must not form an articulation pair.

The charge-dipole bonds $f\cd$ and $f\dc$ decay as $1/r_{12}^2$ at large distances. We decompose therefore $f\cd$ (resp. $f\dc$) into $f\cd(i,j)=f\cd^{\text{sr}}(i,j)+\philr\cd(i,j)$, where $\philr\cd(i,j)=-\beta v\cd^{\text{lr}}(i,j)$ is the electrostatic charge-dipole interaction and $f\cd^{\text{sr}}$ decays more rapidly. The graphs contributing to $w(1,2)$ at large relative distances $R_{12}$ are those which have only a single bond $\philr\cd(1,i)$ attached to the ion $1$, and a single bond $\philr\dc(j,2)$ attached to the ion $2$:
\eq{	\label{w=graphs}
\psfrag{h}{{\footnotesize $h(3,4)$}}
\psfrag{1}{{\footnotesize $1$}}
\psfrag{2}{{\footnotesize $2$}}
\psfrag{3}{{\footnotesize $3$}}
\psfrag{4}{{\footnotesize $4$}}
  -\beta w(1,2) \simeq -\beta V\cc(1,2) + \Graph{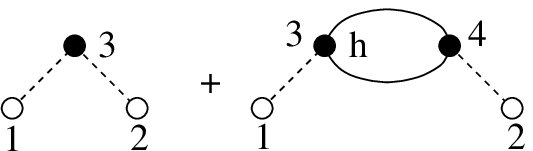}, \qquad R_{12}\to\infty.
}
In the second graph, the solvent molecules $3$ and $4$ are connected by the total (inhomogeneous) dipole-dipole correlation function $h(3,4)$. The contribution of the first graph is
\eq{	\label{convol}
    \int_{z_3>0}\!\romd\ver_3 \int\romd\Omega_3  
   	\, \philr\cd(1,3) \rho(3) \philr\dc(3,2)
	= \Big[ \phi\cd+ \frac{1-\eps'}{1+\eps'}\, \phi\cd^* \Big]
	\star \Big[ \phi\dc+ \frac{1-\eps'}{1+\eps'}\, \phi\dc^* \Big],
}
where we denote the convolution by a star.  If the convolution $\phi\cd  \star \phi\dc$ were perfomed over the whole space (\emph{i.e.} with $\rho(3)$ replaced by a constant), it is clear from dimensional analysis that it would lead to a term decaying as $1/r$ (because we would integrate two bonds decaying as $1/r^2$ over a 3-dimensional space). The regions where the point $3$ is close to the root points leads to terms decaying as $1/r^2$, since $\phi\cd$ and $\phi\dc\sim r^{-2}$. The asymptotic behaviour of the convolution is hence determined by the integration over the regions where the point $3$ is far away from the points 1 and 2.

When focusing on the asymptotic behaviour $R_{12}\to\infty$ of the convolution \eqref{convol}, it is legitimate to replace $\rho(3)$ by the step profile $\Theta(z_3)\rho/(4\pi)$.
Indeed, let $\phi\cd  \star \phi\dc =  q_1 q_2 \beta^2 J(z_1,z_2,R_{12})$ and take the limit $R_{12}\to\infty$ by setting $R_{12}=R'_{12}/\gamma$ with $\gamma \to 0$.
From \eqref{def v_cd}, we have
\begin{multline}	\label{e.20}
   J(z_1,z_2,\frac{R_{12}}{\gamma})    = \gamma
    \int\limits_{z'_3>0}\romd\ver'_3 \int\romd\Omega_3\, \Big(\vp_3\cdot\grad'_3\frac{1}{|\ver'_3|}\Big)
   \Big(\vp_3\cdot\grad'_3 \frac{1}{\Big|\!\Big| \substack{\R'_3-\R'_{12} \\ z'_3-\gamma z_{12}}\Big|\! \Big|} \Big) \rho(\frac{z'_3}{\gamma}+z_1,\vp_3)
\end{multline}
where we introduced the change of variable $\ver_3'=\gamma(\ver_3-\ver_1)$. $J(z_1,z_2,{R_{12}}/{\gamma})$ is hence proportional to $\gamma$ as $\gamma\to 0$ (assuming that $z_{12}$ does not increase faster than $1/\gamma$), so that $J(z_1,z_2,R_{12})$ decays as $1/R_{12}$ at large distances. Moreover, $\lim_{\gamma\to 0}\rho({z'_3}/{\gamma}+z_1,\vp_3)=\rho/(4\pi)$, and the convolutions in eq.~\eqref{convol} can therefore be calculated with the step profile $\Theta(z_3)\rho/(4\pi)$ in place of $\rho(3)$ when $R_{12}\to\infty$. The results for these integrals are shown on Table~I.

\begin{table}[h]
\label{t.1}
\begin{center}
\caption{Convolutions over the region $z>0$ of interaction bonds between dipoles and charges:}
\begin{tabular}{c|@{\hspace{0.2pt}}|c|c|c|c|}
$\nearrow$	& $\phi\dd$ & $\phi\dd^*$ & $\phi\dc$ & $\phi\dc^*$
\\ \hline\cline{1-5} \raisebox{-1.5ex}{\rule{0mm}{4ex}}
$\phi\cd$	& $-3y\phi\cd + \frac{3y}{2}\phi\cd^*$ & $-\frac{3y}{2}\phi\cd^*$ & $-3y\phi\cc + \frac{3y}{2}\phi\cc^*$ & $-\frac{3y}{2}\phi\cc^*$
\\  \cline{1-5}\raisebox{-1.5ex}{\rule{0mm}{4ex}}
$\phi\cd^*$	& $- \frac{3y}{2}\phi\cd^*$ & $-\frac{3y}{2}\phi\cd^*$ & $- \frac{3y}{2}\phi\cc^*$ & $-\frac{3y}{2}\phi\cc^*$\\
\cline{1-5}
\end{tabular}
\vspace{1em}
\end{center}
\end{table}

When the decomposition $c(1,2)=c\sr(1,2)+\phi\dd(1,2)+\phi\dd^*(1,2)$ is introduced in the Ornstein-Zernike equation, the correlation function $h(1,2)$ splits into the short-range part $h\sr(1,2)$ [see eq.~\eqref{OZ h_sr}] and a long-range part $h\lr(1,2)$. Consider the second graph of \eqref{w=graphs} with $h\sr$ in place of $h$ and add this term to eq.~\eqref{convol}:
\begin{equation}	\label{hyper}
    \int_{z_3>0}\!\romd3\romd4
   	\, \philr\cd(1,3) \frac{\rho}{4\pi} \Big[ \delta(3,4)+\frac{\rho}{4\pi} h\sr(3,4) \Big] \philr\dc(4,2), \qquad R_{12}\to\infty.
\end{equation}
Since the points 3 and 4 are close together (because $h\sr(3,4)$ is short-ranged) and far away from the root points, they can be treated as a hypervertex. Using \eqref{def v_cd} and \eqref{eps bulk}, expr.~\eqref{hyper} becomes
\eq{	\label{G1}
   \frac{\eps-1}{3y}     \int_{z_3>0}\!\romd3  
   	\, \philr\cd(1,3) \frac{\rho}{4\pi} \philr\dc(3,2)
	= \frac{\eps-1}{3y} \Big[ \phi\cd+ \frac{1-\eps'}{1+\eps'}\, \phi\cd^* \Big]
	\star \Big[ \phi\dc+ \frac{1-\eps'}{1+\eps'}\, \phi\dc^* \Big].
}
The long-range part $h\lr(3,4)$ has been studied by Badiali~\cite{Bad88} and Badiali and Forstmann~\cite{BadFor}. For points 3 and 4 far apart and far away from the interface, it is given by
\eq{
  h\lr(3,4) \simeq \left(\frac{\eps-1}{3y}\right)^2 \frac{1}{\eps} \Big[ \phi\dd(3,4) + \frac{\eps-\eps'}{\eps+\eps'} \, \phi\dd^*(3,4) \Big].
}
The second graph in \eqref{w=graphs} with $h\lr$ in place of $h$ gives thus the contribution
\eq{	\label{G2}
  \left(\frac{\eps-1}{3y}\right)^2 \frac{1}{\eps}  \left[ \phi\cd+ \frac{1-\eps'}{1+\eps'}\, \phi\cd^* \right] \star \left[ \phi\dd(3,4) + \frac{\eps-\eps'}{\eps+\eps'} \, \phi\dd^*(3,4) \right]
  \star \left[ \phi\dc+ \frac{1-\eps'}{1+\eps'}\, \phi\dc^* \right].
}
According to \eqref{w=graphs}, the potential of mean force is given at large distances by $V\cc$ minus the expressions \eqref{G1}  and \eqref{G2} over $\beta$. Using the results shown on Table~I for the convolutions, we find after some simple algebra:
\eq{
  w(1,2) \simeq  \frac{1}{\eps}  \frac{q_1 q_2}{|\ver_{1}-\ver_2|} + \frac{1}{\eps} \frac{\eps-\eps'}{\eps+\eps'} \, \frac{q_1 q_2}{|\ver_{1}-\ver^*_2|} ,
  \qquad R_{12}\to\infty.
}
The present calculation shows therefore that, even though the ions are close to the interface, their effective interaction at large relative distances along the wall is given by the standard formula of macroscopic electrostatics. Local effective dielectric functions for solvated macromolecules obtained from continuum electrostatics models \cite{Honig} have thus some microscopic basis.





\acknowledgments
V.B. thanks the Royal Society and the Swiss National Science Foundation for financial support.

\end{document}